\documentclass[11pt]{article}
\usepackage[T1]{fontenc}
\usepackage{lmodern}
\usepackage[utf8]{inputenc}
\usepackage[a4paper,margin=1in]{geometry}
\usepackage{graphicx}
\usepackage{multirow}
\usepackage{hyperref}
\usepackage{listings}
\usepackage{amsmath,amssymb,amsthm}

\newcommand{\thedate}{July 27, 2020}
\newcommand{\fundingnote}{The research has been supported by the European Union, co-financed by the European Social Fund. (EFOP-3.6.2-16-2017-00013).}
\newcommand{\mscclass}{68N15, 68Q55, 68Q65}
\newcommand{\acmclass}{D.3.4 [\textbf{Processors}]: Interpreters; F.3.2 [\textbf{Semantics of Programming Languages}]: Program analysis and translation}
\newcommand{\paperkeywords}{emulation-completeness, self-emulation, language expressiveness, semantic equivalence, interpreter design, runtime observability}

\newtheorem{definition}{Definition}[section]

\title{Emulation-Completeness of Programming Languages}
\author{Gregory Morse \and
Tam\'as Kozsik}
\date{\thedate}

\begin{document}

\maketitle

\begin{abstract}
We study when a programming language can emulate programs written in that same language without delegating the guest program back to the host evaluator or compiler.  We call this property \emph{emulation-completeness}.  The central observation is that Turing-completeness by itself is not enough: a self-emulator must not only compute the guest program's result, but must also account for the guest-visible state on which realistic programs depend, including control flow, exceptions, callbacks, timing, memory usage, and runtime metadata such as stack traces or line numbers.

This paper is a systematization paper.  Its contribution is not a new emulator implementation, but a precise vocabulary and a structured taxonomy for reasoning about self-emulation.  We distinguish source-level evaluation from compiled-code emulation, define syntactic and compiled-code emulation-completeness, and separate \emph{weak} from \emph{strong} emulation-completeness according to how much observable runtime behavior must be preserved.  We then organize the requirements into two classes: language-side requirements, which determine whether the guest semantics can be represented explicitly inside the language, and emulator-side requirements, which determine whether the resulting emulator can faithfully mask or reproduce relevant observations.

The discussion is grounded by concrete examples, including publicly documented details from Erlang, where argument limits, bitstring pattern matching, and message reception expose subtle mismatches between direct execution and self-emulation.  The resulting framework is intended as guidance for language designers, implementers of evaluators and emulators, and researchers interested in secure sandboxing, decompilation, and reflective execution.
\end{abstract}

\begin{center}
\small Department of Programming Languages and Compilers, Faculty of Informatics, ELTE, E\"{o}tv\"{o}s Lor\'and University, 1/C P\'azm\'any P\'eter s\'et\'any, Budapest, 1117, Hungary\\
\texttt{morse@inf.elte.hu} \quad \texttt{kto@elte.hu}
\end{center}

\noindent\textbf{MSC 2010:} \mscclass

\noindent\textbf{ACM CCS:} \acmclass


\noindent\textbf{Funding:} \fundingnote

\medskip

\section{Introduction}
Emulation is usually discussed at the machine level: an emulator reproduces the behavior of a processor, an operating system, or a virtual machine.  In practice, such systems are typically implemented in low-level languages because performance and precise control over runtime state are essential.  Yet there is a closely related language-design question that is much less studied: to what extent can a programming language emulate programs written in that same language?

Most languages do not provide direct syntactic support for exact self-emulation of their own compiled artifacts, and many do not even make the relevant runtime metadata available in a form that can be reconstructed faithfully.  This gap is one reason the topic is easy to underestimate: the core computational task may be expressible, while the surrounding semantic obligations are not.

At first sight, the answer may appear to be immediate.  If a language is Turing-complete, then it can in principle encode arbitrary computations, and therefore it should be able to encode an emulator as well.  This intuition is directionally correct but incomplete.  A realistic emulator does not merely compute a final value.  It must decode guest constructs, maintain explicit guest state, preserve the guest program's exception behavior, interact coherently with callbacks and libraries, and, in the strongest setting, hide or reproduce observations about timing, memory, stack traces, and other runtime artifacts.  These obligations go substantially beyond bare computability.

The goal of this paper is therefore to give a precise and usable formulation of \emph{emulation-completeness}.  We are interested in the following question: when can a language express a self-emulator that simulates its own programs by fetching, decoding, and advancing an explicit guest state, rather than by delegating execution back to a built-in \texttt{eval}, JIT compiler, or byte-code loader?  The distinction matters whenever self-emulation is used as part of security instrumentation, testing, symbolic evaluation, decompilation, or reflective tooling.

This paper makes four contributions.
\begin{enumerate}
\item It defines a clearer terminology for self-emulation, distinguishing evaluation, execution, and emulation, and separating source-level and compiled-code settings.
\item It introduces a layered notion of emulation-completeness, including syntactic versus compiled-code emulation-completeness and weak versus strong emulation-completeness.
\item It presents a taxonomy of language-side and emulator-side requirements that must be checked when assessing whether a language can faithfully emulate itself.
\item It grounds the taxonomy with concrete examples, especially from Erlang, where publicly visible implementation choices reveal how seemingly small language details can obstruct faithful self-emulation.
\end{enumerate}

Equally importantly, we also clarify the scope of the contribution.  This is a systematization paper, not a completeness theorem for a particular industrial language and not a performance study of a specific emulator.  The purpose is to make the research question precise and to expose the technical obligations that any positive or negative result would need to address.

Motivations for studying emulation-completeness include security analysis, sandboxing, controlled execution of untrusted code, testing of critical systems, symbolic evaluation, and decompilation.  These are exactly the settings in which a mismatch between direct execution and emulated execution is not merely an implementation nuisance but a semantic issue.  In particular, anti-emulation and anti-sandbox techniques show that observable differences in timing, memory footprint, or fault behavior can become semantically relevant to the guest program itself \cite{10.1007/978-3-540-75496-1_1,10.1007/978-3-642-23644-0_18}.

\section{Background and Definitions}

\subsection{Evaluation, execution, and emulation}
We use \emph{evaluation} for direct traversal of a source-level representation such as an abstract syntax tree, \emph{execution} for running compiled code on the target runtime or hardware, and \emph{emulation} for simulation of guest behavior by an explicit interpreter state that is itself implemented in the host language.  This paper focuses on the third case.

The distinction is important because a built-in \texttt{eval} is not, by itself, evidence of emulation-completeness.  If a host program receives a guest program representation and simply delegates it back to the host compiler or evaluator, then the host language is using its ordinary execution path rather than explicitly representing the guest state.  Such delegation may be useful in practice, but it avoids the expressive question studied here.

Table \ref{tbl:emulation} summarizes the terminology used throughout the paper.

\begin{table*}[!htb]
\centering
\caption{Execution contexts considered in this paper}
\begin{tabular}{l | l | l | l | l}
& \textbf{Emulator} & \textbf{VM} & \textbf{JIT VM} & \textbf{Hardware}\\ \hline
\textbf{Syntax tree} & Evaluation & Evaluation & On-the-fly compilation & --\\ \hline
\textbf{Byte-code} & Emulation & Execution & $\rightarrow$ execution & Execution
\end{tabular}
\label{tbl:emulation}
\end{table*}

\subsection{A running example}
To keep the discussion concrete, consider a language with higher-order functions, exceptions, pattern matching, concurrency, and reflective operations such as stack-trace inspection or time queries.  A self-emulator for this language receives a representation of a guest program together with an explicit guest state containing, at minimum, environments, a store, a control stack, and any language-specific runtime metadata.  For each guest step, the emulator fetches the next guest construct, decodes it, updates the guest state, and only then continues.

This running example already exposes the two main themes of the paper.  First, the host language must be expressive enough to represent each guest transition without silently delegating the difficult cases to the host runtime.  Second, the resulting emulator must control what the guest can observe, because reflective operations can reveal the difference between guest execution and host execution.  The later sections analyze these obligations systematically.

\subsection{Program representations and observables}
Source code, abstract syntax trees, and byte-code are all representations of programs, but they expose different structure.  An abstract syntax tree normally preserves the nesting structure of the source language; byte-code is usually flatter, lower-level, and closer to a machine model.  This difference is relevant because some languages can readily evaluate their own syntax trees while still lacking a faithful route for emulating their own compiled code.

We also need a notion of \emph{observable behavior}.  For the purposes of this paper, the observables of a guest program include its final value, uncaught exceptions, I/O behavior, interactions with callbacks and the host environment, and any information that the language exposes through reflection or introspection, such as stack traces, line numbers, timing, or memory availability.  The precise set of observables is language-dependent, but the taxonomy below is driven by exactly these observations.

Semantic equivalence can therefore be defined at different granularities depending on which effects are considered observable.  If timing is included at full precision, then the only perfectly equivalent implementation may be the original source or byte-code itself.  In practice, however, languages provide only partial timing guarantees, and developers routinely expect the same program to run on very different machines and runtimes.  A useful working notion of equivalence is thus usually inherited from the compiler and runtime implementation: the space of emitted normal and optimized executions that programmers understand as ``the same program.''  This notion is not fully formal, but it is exactly the level at which self-emulation is judged in practice.

\begin{definition}[Self-emulator]
Let $L$ be a programming language and let $\mathit{repr}_L$ encode programs of $L$ as ordinary values of $L$.  A \emph{self-emulator} for $L$ is a program $U_L$ written in $L$ such that, for every guest program $p$ in the chosen program class, $U_L(\mathit{repr}_L(p), s_0)$ reproduces the guest-visible behavior of $p$ by maintaining an explicit guest state $s$ and advancing that state step by step.  In particular, the guest control flow is represented inside $U_L$; it is not simply delegated to the host evaluator or compiler.
\end{definition}

\begin{definition}[Syntactic and compiled-code emulation-completeness]
A language is \emph{syntactically emulation-complete} if it admits a self-emulator for its own source-level representation.  It is \emph{compiled-code emulation-complete} if it admits a self-emulator for its own compiled representation as well.  The second property is typically stronger because compilation often introduces lower-level operations, optimizations, and metadata that the source language does not expose directly.
\end{definition}

\begin{definition}[Weak and strong emulation-completeness]
\emph{Weak emulation-completeness} requires preservation of guest-visible behavior under a fixed or appropriately synchronized external state.  \emph{Strong emulation-completeness} additionally requires that every observation available to the guest---including observations derived from timing, memory usage, reflection, or fault behavior---remain indistinguishable from direct execution.  Strong emulation-completeness is therefore primarily a theoretical ideal.
\end{definition}

One way to sharpen this distinction is to imagine two abstract machines: one executing the guest program directly after translation to a machine-compatible form, and one executing the emulator together with a representation of the guest.  Both machines may interact with an external state area representing clocks, devices, or other environmental inputs.  Strong emulation-completeness requires indistinguishability even as that external state evolves during execution; weak emulation-completeness assumes that the external state is fixed or synchronized closely enough that the guest cannot exploit the difference.

These definitions explain why Turing-completeness is insufficient.  Computability guarantees that some simulation exists in principle, but emulation-completeness asks whether the language and runtime expose enough structure to realize that simulation faithfully under the observations that matter.

There are also practical reasons why this distinction matters.  Read-evaluate-print loops and \texttt{eval}-style mechanisms are common, and some languages implement substantial parts of these facilities within the language itself, as in Erlang \cite{Armstrong:2007:HE:1238844.1238850}.  Self-emulation is also relevant to sandboxing, as shown by JavaScript-in-JavaScript systems \cite{181092}.  It can additionally serve as a protection wrapper against debugging or tracing, since the guest code executes through a virtualized layer of interpretation rather than directly on the host path.  Finally, in symbolic evaluation or decompilation, exact agreement between explicit interpretation and normal execution is often essential.

\section{Language Capabilities}
We first consider requirements imposed by the language itself.  These are the properties that determine whether guest execution can be represented explicitly as data and state transitions inside the host language.  If any of these requirements fail, then even a carefully engineered emulator may be forced to appeal to external mechanisms or to silently change the guest semantics.

\subsection{Fundamentals}
At a minimum, the language must be able to fetch guest instructions or syntax, decode them, and maintain an explicit guest state.  In most practical languages this already presupposes familiar building blocks such as data structures, control flow, arithmetic, and indirect references.  Strictly speaking, Turing-completeness is not the formal criterion here, but without a sufficiently rich notion of state and control, faithful self-emulation is unlikely.

The critical issue is not whether the language can compute the same mathematical function as the guest program.  Rather, it is whether each guest construct can be represented in the emulator as a controlled transition over explicit data.  This is why low-level intermediate languages or byte-code formats may sometimes be easier to emulate than their higher-level source languages: the former often expose smaller and more uniform transition steps.

For compiled languages, an additional complication is that the byte-code format is often undocumented, unstable, or treated as an implementation detail.  Such a format may be closer to emulation-completeness than the surface language while still being unsuitable as a long-term semantic target because it is not a maintained language in its own right.

\subsection{Function arguments}
Higher-order functions create a subtle but important requirement.  Whenever a function value is passed to the environment or returned to code that may later invoke it, the emulator must regain control before the guest body is executed.  Operationally, this means that callbacks need to be wrapped with an additional layer that re-enters the emulator and restores the relevant guest state.

That wrapper often needs extra information: the current guest state, or at least a unique handle with which the guest state can be recovered.  If the language imposes a hard limit on function arguments, closure captures, or callable representations, then perfectly ordinary guest programs may become unemulatable because the wrapper has nowhere to store the extra control information.  Languages that support variadic calling conventions, argument lists, or stable runtime identifiers for function values make this problem much easier.

This issue is not merely hypothetical.  In Erlang's \lstinline|erl_eval| implementation, the authors explicitly work around argument-handling limits in evaluator wrappers and comment that the solution is ``a really ugly hack'' \cite{ErlStdLibEval}.  The example is instructive because the guest language itself supports programs whose direct compiled execution does not suffer from the evaluator's additional restriction.  The mismatch therefore comes from the self-emulation route, not from the source language alone.

\subsection{Compound and complex language constructs}
Some language features are not atomic.  Pattern matching, message reception, exception propagation, and other rich constructs may expand into several lower-level steps or may depend on runtime services that are only partially visible at the source level.  If the emulator cannot express these steps directly, it may be forced to approximate the construct using extra exceptions, auxiliary control flow, or an \texttt{eval}-like escape hatch.  Once that happens, semantic fidelity becomes questionable.

The requirement can be stated simply: either the language must expose the constituent operations of the construct, or it must let the emulator represent the whole construct as explicit guest state and manipulate it without executing host behavior that the guest did not request.  In particular, resorting to extra exception handling, host-side compilation, on-the-fly code generation, or another \texttt{eval}-style mechanism is not a faithful solution if it inserts control flow or effects that the guest program would not otherwise traverse.  The problem becomes particularly visible when a source-level construct corresponds to a delicate compiled-code protocol.

Erlang again offers useful examples.  In the \lstinline|eval_bits| code referenced from \lstinline|erl_eval|, bitstring pattern matching is effectively decomposed into smaller steps whose failure is detected with \lstinline|catch| and re-packaged as a different outcome \cite{ErlStdLibBitEval}.  This is convenient engineering, but it introduces exception behavior that is not identical to the compiled path.  Similarly, the \lstinline|receive| construct is handled in the \lstinline|prim_eval| code from the runtime system rather than purely within the language \cite{ErlErtsRcvEval}.  The public documentation for \lstinline|bump_reductions| even warns that the interface is unstable and implementation-specific \cite{ErlErlangManual}.  These examples show that apparently local implementation shortcuts expose genuine language-design pressure points.

\section{Emulator Capabilities}
Language expressiveness is only half of the problem.  Even if every guest step can be represented inside the host language, the resulting emulator must still mediate the guest's interaction with the runtime and environment.  In practice, strong self-emulation fails not because the next arithmetic operation is hard to encode, but because the guest can observe some difference between host execution and emulated execution.

\subsection{System and emulator state introspection}
Many runtimes expose direct or indirect access to system state.  Programs may query the clock, the amount of available memory, the current process identifier, the structure of a stack trace, the version of the runtime, or other metadata.  These observables are problematic because the emulator itself perturbs them.

Some observables are easy to identify but hard to reproduce; others are hard even to enumerate.  Direct reflection on emulator-owned data is usually easier to control than indirect observations of shared system state.  In large software stacks, even identifying every relevant source of observability can be a maintenance problem, because the surrounding system may evolve faster than the emulator can be audited.  Table \ref{tbl:introspect} summarizes this asymmetry.

\begin{table*}[!htb]
\centering
\caption{Difficulty of masking observable differences}
\begin{tabular}{l | l | l}
& \textbf{System state} & \textbf{Emulator state}\\ \hline
\textbf{Direct observation} & Harder & Easier\\ \hline
\textbf{Indirect observation} & Hardest & Hard
\end{tabular}
\label{tbl:introspect}
\end{table*}

\subsection{Timing}
Timing is among the clearest reasons why strong emulation-completeness is mainly a theoretical ideal.  A faithful emulator consumes host computation.  Unless the surrounding environment is specially engineered, that extra work changes the elapsed time seen by the guest.  Modern runtimes also expose several distinct notions of time: wall-clock time, process time, per-thread CPU usage, timer interrupts, and interleaving behavior in concurrent programs.

Wall-clock time is especially unforgiving.  A guest that consults UTC or communicates with an external system can compare the progress of the emulated execution with time outside the host process.  No purely internal trick can make an arbitrarily expensive emulator appear to have taken no time at all.  This is one of the strongest arguments for treating strong emulation-completeness as a limiting notion rather than a practical engineering target.  By contrast, internal clocks derived from host counters may sometimes be virtualized or offset.  The crucial difficulty is that universal time and remote interaction are not merely local runtime artifacts; they are part of a broader environment that the emulator does not control.

Timing also affects parallel code.  Deadlocks, races, starvation, and scheduling-sensitive bugs may appear or disappear when the emulator perturbs the relative ordering of events.  If the language supports callbacks, threads, processes, or lightweight actors, then the emulator must track per-task guest state and re-enter the correct state whenever the environment calls back into guest code.  Even seemingly secondary metrics such as per-process or per-thread CPU usage can then become indirect evidence of emulation.

A further complication comes from compiled-code optimizations.  Byte-code or VM implementations may use specialized opcodes or optimized fast paths that collapse several source-level steps into one.  A source-level emulator that reconstructs the same result may still disagree on timing or intermediate observables.  In principle, the host language might expose internal low-level operations that let the emulator imitate the optimized path directly, but such interfaces tend to be fragile and unmaintainable.  Table \ref{tbl:timingdisc} summarizes the main timing-related discrepancies.

At the limit, a strong solution would require modeling the underlying runtime closely enough to simulate the guest clock and synchronize guest-visible steps across concurrent tasks.  On proprietary hardware or opaque virtual machines this is generally infeasible without detailed specifications or extensive empirical reverse engineering, which again explains why the strong notion is better understood as a theoretical bound.

\begin{table*}[!htb]
\centering
\caption{Timing discrepancies and typical responses}
\begin{tabular}{p{0.44\textwidth} | p{0.44\textwidth}}
\textbf{Category} & \textbf{Typical response}\\ \hline
Internal time observable via system inspection & Hook or virtualize every direct and indirect clock access\\ \hline
Time derived from emulator-owned state & Keep a separate guest clock and expose only that clock\\ \hline
External time (for example UTC or remote interaction) & Requires a contrived or specially synchronized environment\\ \hline
Inter-task timing in parallel code & Simulate scheduling and synchronize guest-visible steps\\ \hline
Timing differences caused by optimized byte-code paths & Model the relevant optimization or weaken the equivalence criterion
\end{tabular}
\label{tbl:timingdisc}
\end{table*}

\subsection{Memory}
Memory creates an analogous set of problems.  The emulator consumes heap space, stack space, and allocator metadata that the directly executed guest program would not consume in the same way.  If the guest can observe memory pressure, then emulation changes the program's environment.

There are two distinct issues.  The first is absolute limits: an emulator may cause an out-of-memory condition or stack overflow earlier than direct execution would.  The second is introspection: some languages or runtimes expose used memory, remaining memory, allocator behavior, or recursion depth.  Even when these metrics are only approximate, they may still be sufficient to detect emulation.

One partial mitigation is to keep the emulator's control structures off the host call stack, using explicit heap-allocated stacks instead of deep host recursion.  This does not eliminate the problem, but it avoids one common source of divergence.  More fundamentally, any host memory consumed by the emulator is already a discrepancy: the problem is not merely measurability, but the fact that the emulator consumes part of a finite global budget that the directly executed guest could otherwise have used.  In practice, memory-based detection is usually less robust than timing-based detection, but it cannot be ignored in a strong account.  Table \ref{tbl:memorydisc} summarizes the main cases.

\begin{table*}[!htb]
\centering
\caption{Memory discrepancies and typical responses}
\begin{tabular}{p{0.44\textwidth} | p{0.44\textwidth}}
\textbf{Category} & \textbf{Typical response}\\ \hline
System-wide out-of-memory limits & Requires a contrived bound or a conservative resource budget\\ \hline
Emulator stack growth & Use an explicit heap-allocated guest stack\\ \hline
Memory observable via system inspection & Hook or virtualize the relevant inspection interfaces\\ \hline
Memory derived from emulator-owned state & Hide the host footprint and expose only guest-accounted usage
\end{tabular}
\label{tbl:memorydisc}
\end{table*}

\subsection{Other system interactions}
An emulator is not a full virtual machine unless it also virtualizes the surrounding operating environment.  In the setting studied here, system calls, I/O, and foreign-library boundaries remain real interactions with the host system.  This is both a burden and an opportunity: those boundaries are precisely where the emulator can interpose and translate between guest semantics and host behavior.

The boundary must be defined clearly.  If application libraries are written in the same language and available in the same representation as the guest program, then they should ordinarily be considered part of the emulated world.  If they are available only in opaque compiled form, then for the purposes of source-level emulation they behave more like foreign system interfaces.  This distinction matters because otherwise the emulator may quietly rely on compiled components that bypass the explicit guest state.

The emulator should also minimize its own use of scarce or measurable resources.  File descriptors, kernel objects, and process identifiers can become observables in exactly the same way as memory and time.

\subsection{Callback functions}
Callbacks are especially difficult because control returns from the host environment into guest code at points not necessarily visible in the original guest syntax.  To remain faithful, the emulator must intercept the creation of callable guest values or, failing that, identify every place where the environment may later call them.

If the language offers stable runtime identities for function values, explicit wrapping points, or metadata that marks callback-capable constructs, then the problem is manageable.  Without such support, the emulator is forced into brittle whole-system analysis.  Enumerating all callback sources in the surrounding platform is not really a property of the language any more, but in practice it becomes part of the engineering burden of self-emulation.

Generated functions make the issue harder still.  A static declaration of possible callbacks is rarely sufficient when function values are synthesized at run time.  It may seem that a sufficiently clever emulator could simply interpret all data semantically and recognize function pointers on the fly, but that pushes the problem beyond emulation and into whole-program semantic recovery.  In von Neumann-style settings, code and data remain ambiguous until a control transfer actually occurs, and even segmented memory models do not eliminate the issue completely.  This is another example of how a language feature that is benign under direct execution can become a first-class obstacle under self-emulation.

\subsection{Overlooked and hidden state information}
The most treacherous state is often the state that is not initially documented as part of the language semantics.  Modern hardware and runtimes expose behavior through caches, speculative execution, scheduler metadata, debug information, and optimization artifacts.  Even if such details are only indirectly observable, they can still matter when the guest program is trying to distinguish emulated execution from direct execution.  The broad lesson of vulnerabilities such as Spectre and Meltdown is that microarchitectural state can become semantically relevant once timing or other side channels make it observable.

\subsubsection{Source metadata in compiled code}
Line numbers, source locations, symbol tables, and even complete fragments of source code may be stored in byte-code files or auxiliary debugging information.  If the language exposes any of this information at run time, then the emulator must decide whether the metadata belongs to the guest state and, if so, how it is reproduced faithfully.

\subsubsection{Stack traces}
Stack traces are a particularly important form of metadata because they are often produced automatically by the runtime rather than by user code.  A self-emulator that simply throws a host exception will typically expose the emulator's own control stack instead of the guest stack.  Therefore the guest stack must be represented explicitly and reified whenever the language exposes it.

\section{Differences Between Syntax-Tree Evaluation and Compiled-Code Execution}
Even when a language readily supports source-level evaluation, that does not imply agreement with compiled execution.  The difference is not merely one of performance.  Compilation may enforce static restrictions, normalize control flow, or emit specialized instructions whose observable behavior differs from a straightforward traversal of the syntax tree.

\subsection{Evaluation level}
Languages support evaluation at different granularities: an expression, a declaration, a function body, or an entire module.  The chosen granularity affects observable behavior.  For example, evaluating a single expression may not create the same line-number information, stack frames, or module initialization effects as compiling and executing the enclosing unit.

\subsection{The syntax tree may be easier to evaluate than the compiled code}
Compilers commonly perform semantic checks after parsing.  Scope resolution, binding analysis, exhaustiveness checks, or definite-assignment checks may reject programs that are nevertheless representable as syntax trees.  A direct syntax-tree evaluator that bypasses these checks does not implement the same language as the compiler any more; it implements a more permissive variant.

\subsubsection{Variable assignment and scoping}
This discrepancy is easiest to see with scoping rules.  A syntax-tree evaluator may be able to begin traversing code immediately, but the compiled language may reject that same code because a variable is out of scope, used before definition, or bound inconsistently across control-flow paths.  If the goal is semantic equivalence to compiled execution, then the evaluator must replicate the relevant compiler checks before execution starts.

\subsection{Variable emission fence}
Compilation also introduces structural commitments that are not present explicitly in the source syntax.  We use the term \emph{variable emission fence} for a program point at which variables, closures, or unique runtime identities must already have been materialized before execution proceeds.  Lambda capture is the simplest example: even if the source syntax looks declarative, compiled execution usually commits to a closure object at a definite point.

If the emulator attaches unique identifiers to guest functions, closures, or temporaries, then these identifiers also create fences because the act of emitting them is observable through later equality checks, debugging information, or serialization.  Table \ref{tbl:varemitfence} lists representative cases.

\begin{table*}[!htb]
\centering
\caption{Representative variable-emission fences}
\begin{tabular}{p{0.44\textwidth} | p{0.44\textwidth}}
\textbf{Context} & \textbf{Reason}\\ \hline
Block or lexical scope boundary & Variables must respect the compiled scoping discipline\\ \hline
Anonymous or generated function value & A stable runtime identity may become observable\\ \hline
Closure capture & Captured variables must be materialized before later use
\end{tabular}
\label{tbl:varemitfence}
\end{table*}

\section{Bugs, Faults, and Logic Errors}
If the underlying hardware, runtime, or virtual machine contains a bug, then a faithful emulator must decide whether that bug is part of the guest semantics.  In strong emulation-completeness the answer is effectively yes whenever the guest can observe the bug.  Correcting the bug inside the emulator may inadvertently make the emulator easier to detect than the original execution environment.

The same applies to fault behavior.  If the guest program would raise an uncaught exception, crash, or enter an infinite loop under direct execution, then the emulator should reproduce the same guest-visible behavior.  This includes exception handlers, stack traces, crash metadata, and any persistent diagnostic artifacts that later executions might inspect.  In particular, if an uncaught fault exposes the host emulator stack or leaves a crash footprint characteristic of the emulator rather than the guest, then the emulation has already leaked information.  For a single run, some of these issues may appear peripheral; for repeated executions or adversarial workloads, they are part of the observable semantics.

Likewise, logic errors do not disappear merely because execution is emulated.  A guest program that spins forever and consumes all available CPU should remain capable of doing so under emulation if the goal is faithful behavior.  In that sense, the emulator is not a repair mechanism; it is another execution context whose task is to preserve, not sanitize, the original behavior.

\section{Emulator Detection}
Anti-emulation techniques are a useful reality check for the definitions above.  If a guest program can measure timing precisely enough, inspect caches or CPU anomalies, or infer the presence of instrumentation through resource usage, then a self-emulator that is extensionally correct on final outputs may still fail dramatically as an emulator.

Raffetseder, Kruegel, and Kirda identify several ways in which system emulators can be detected, including instruction-set anomalies, caching behavior, and timing differences \cite{10.1007/978-3-540-75496-1_1}.  Lindorfer, Kolbitsch, and Milani Comparetti show a similar arms race in malware sandboxes, where environment-sensitive samples detect the analysis setting and change behavior accordingly \cite{10.1007/978-3-642-23644-0_18}.  These works concern machine-level environments, but the underlying lesson transfers directly: once the guest can observe the host environment, the fidelity of the emulator becomes a semantic property, not just a performance property.

\section{Summary}
The discussion above separates emulation-completeness into a collection of checkable obligations.  Tables \ref{tbl:langlist} and \ref{tbl:emulist} condense the resulting taxonomy into a language-oriented checklist and an emulator-oriented checklist.

\begin{table*}[!htb]
\centering
\caption{Language-oriented checklist for emulation-completeness}
\begin{tabular}{p{0.44\textwidth} | p{0.44\textwidth}}
\textbf{Item} & \textbf{Goal}\\ \hline
Fetch, decode, and maintain guest state & Expressible inside the language itself\\ \hline
All statements and expressions & Reconstructible through explicit data and indirect references\\ \hline
Function arguments and callable values & Unlimited, list-based, or otherwise wrapper-friendly\\ \hline
Function identity & Stable runtime identifier available when needed\\ \hline
Compound or complex operations & Expressible as primitive steps or explicit structured guest state\\ \hline
Source-level evaluation versus compilation & Relevant compiler checks reproducible when semantic equivalence requires them
\end{tabular}
\label{tbl:langlist}
\end{table*}

\begin{table*}[!htb]
\centering
\caption{Emulator-oriented checklist for emulation-completeness}
\begin{tabular}{p{0.44\textwidth} | p{0.44\textwidth}}
\textbf{Item} & \textbf{Goal}\\ \hline
Timing & Perfect synchronization only in the strong setting\\ \hline
Memory boundaries & Host and guest limits made observationally consistent\\ \hline
Reflective and introspection functions & Hooked, virtualized, or explicitly modeled\\ \hline
Memory footprint and resource usage & Hidden or accounted for in guest-visible form\\ \hline
Boundary of emulation & Defined clearly across system calls, libraries, and runtime services\\ \hline
Callbacks and function pointers & Intercepted at creation or reliably identified later\\ \hline
Overlooked or hidden state information & Identified and modeled explicitly\\ \hline
Source metadata and stack traces & Reconstructed as guest artifacts rather than host artifacts
\end{tabular}
\label{tbl:emulist}
\end{table*}

The checklists are not claimed to be mathematically exhaustive.  Rather, they collect the obligations that repeatedly arise in practical attempts to build faithful evaluators, self-emulators, or decompilation-oriented semantic models.  They are informed in part by experience with precise emulation and decompilation based on exact semantic equivalence of byte-code, where many of these issues surface not as philosophical curiosities but as concrete implementation obstacles.  Their main value is to make hidden assumptions explicit.  In particular, they show why strong emulation-completeness tends to require a contrived execution environment in which external observables are synchronized or bounded by design.

\section{Conclusion}
The main thesis of this paper is that self-emulation is a richer expressiveness question than computability alone.  A language may be powerful enough to encode an interpreter and yet still fall short of emulation-completeness because some guest-visible behavior cannot be represented, intercepted, or hidden faithfully.  Conversely, implementation techniques that appear to be mere engineering details---for example, callback wrappers, stack-trace generation, or access to timing APIs---turn out to be central to the semantic question.

By separating language-side requirements from emulator-side requirements, and by distinguishing syntactic from compiled-code emulation as well as weak from strong emulation-completeness, the paper provides a clearer vocabulary for future work.  This vocabulary is useful for language designers who want reflective or sandboxed execution to be principled, for implementers who need to understand why an evaluator diverges from compiled behavior, and for researchers studying secure analysis, virtualization, symbolic execution, or decompilation.

At a practical level, many deployed emulators are built to accomplish a specific task and therefore quite reasonably ignore edge cases that do not matter for that task.  The point of the present taxonomy is not to criticize such systems, but to separate deliberate engineering trade-offs from genuine limitations of the host language or runtime.  Once those trade-offs are made explicit, the research question becomes much clearer.

Several directions remain open.  A natural next step is to instantiate the checklist for specific mainstream languages and to prove sharper positive or negative results for each of them.  Another is to investigate language designs that expose precisely the reflective hooks required for faithful self-emulation without sacrificing maintainability.  Finally, the strongest form of the problem suggests hardware and runtime support questions: if precise timing and resource observability are part of the guest semantics, then some aspects of strong emulation-completeness may need to be designed jointly across language, runtime, and platform.

%
%
%
\bibliographystyle{plain}
\bibliography{biblio}

\end{document}